\documentclass[preprint,showpacs,preprintnumbers,amsmath,amssymb]{revtex4}

\usepackage{graphicx}
\usepackage{dcolumn}
\usepackage{bm}
\usepackage{feynmf}

\begin{document}

\preprint{MTA-PHYS-0503}

\title{Nonspectator effects in $B\to K^*\gamma$ within the vector quark model.  }

\author{Mohammad Ahmady}
 \email{mahmady@mta.ca}
 \homepage{http://www.mta.ca/~mahmady}
\affiliation{Department of Physics, Mount Allison University, 67 York Street, Sackville, New Brunswick,
 Canada E4L 1E6
}%

\author{Farrukh Chishtie}
 \email{fchishti@uwo.ca}
\affiliation{ Department of Applied Mathematics, University of Western Ontario, London, Ontario, Canada N6A 5B7}%

\date{\today}

\begin{abstract}
The tree level FCNC due to the presence of an additional generation of vector quarks result in the leading order
nonspectator contributions to rare $B\to K^*\gamma$ decay mode. These tree level contributions are sensitive
only to $b\to s$ nonunitary parameter $U^{sb}$ and therefore, provide a  direct constraint on this model
parameter.  We obtain the isospin asymmetry between $\bar B^0\to\bar K^{*0}\gamma$ and $B^-\to K^{*-}\gamma$ to
be $\Delta_{0-}=-0.03*\Re (\frac{U^{sb}}{a^c_7V_{tb}V^*_{ts}})$ and the direct CP asymmetry between $B^+\to
K^{*+}\gamma$ and $B^-\to K^{*-}\gamma$ to be
$A_{CP}^{VQM}=0.27\left\vert\frac{U^{sb}}{V_{tb}V_{ts}^*}\right\vert\sin{\theta}\sin{\phi_s}$, where $\theta$ is
the weak phase of $U^{sb}$ and $\phi_s$ is the strong phase of decay amplitude.  We predict a direct CP
asymmetry of around a few percent if the current experimental difference between $\Delta_{0-}$ and $\Delta_{0+}$
is to be explained by the presence of the additional vector quarks.
\end{abstract}
\maketitle

\section{\label{section1}Introduction}

The precision measurement of the radiative decay mode $B\to K^*\gamma$ has provided an exciting opportunity to
test the Standard Model(SM) and beyond. Besides the branching ratio, the isospin asymmetry in this process which
is defined as:
\begin{equation}
\Delta_{0-}=\frac{\Gamma (\bar B^0\to\bar K^{*0}\gamma ) -\Gamma (B^-\to K^{*-}\gamma )}{\Gamma (\bar B^0\to\bar
K^{*0}\gamma )+\Gamma (B^-\to K^{*-}\gamma )}\;\; , \label{isospinasym}
\end{equation}
could prove to be an important observable for examining the SM as well as discriminating between various new
physics scenarios. The data from Belle\cite{belle} and Babar\cite{babar} point to isospin asymmetries of at most
a few percent and consistent with zero within the experimental error:
\begin{eqnarray}
\Delta_{0-}&=& +0.051\pm 0.044({\rm stat.})\;\pm 0.023({\rm sys.})\;\pm 0.024(R^{+/0})\;\;\; (Babar)\; , \label{babar}\\
\Delta_{0+}&=& +0.012\pm 0.044({\rm stat.})\;\pm 0.026({\rm sys.})\;\;\; (Belle)\; , \label{belle}
\end{eqnarray}
 where $\Delta_{0+}$ is defined as in eq.(\ref{isospinasym}) but using the charge conjugate modes.  The last error
 in eq.(\ref{babar}) is due to the uncertainty in the ratio of the branching fractions of the neutral and charged
 B meson production in $\Upsilon (4S)$ decays. This asymmetry is due to the
non-spectator contributions and has been estimated to be around a few percent in the SM within the QCD factorization approach in Refs. \cite{kn} and \cite{bb},
Brodsky-Lepage formalism \cite{petrov} and the perturbative QCD method in Ref. \cite{kms}.  The more accurate measurement of the isospin asymmetry in the near
future and a better understanding of the SM prediction for this observable should provide a sensitive testing venue for possible models of new physics. One such
model is the extension of the SM with an extra generation of iso-singlet quarks\cite{ahmadynagashima}. Unlike the three generations of ordinary quarks in the SM,
both the left- and the right-handed components of the quarks of this additional generation are invariant under $SU(2)_L$ gauge group. Therefore, the flavor
changing weak interactions of these exotic quarks proceeds only through mixing with ordinary quarks and this results in the non-unitarity of the extended
$4\times 4$ quark mixing matrix and thus non-vanishing flavor changing neutral currents (FCNC) at the tree level.  Isospin asymmetry in $B\to K^*\gamma$
transitions offers an excellent physical observable for constraining the parameters of this so-called vector quark model (VQM).  As is shown in our result,
nonspectator effects like the isospin and direct CP asymmetry in $B\to K^*\gamma$ offer the advantage of being sensitive to only one model parameter, namely the
non-unitarity parameter $U^{sb}$ and therefore, can provide a good constraint on the size of the FCNC in the context of VQM irrespective of the masses of the
additional quarks.

\section{\label{section2}Isospin symmetry breaking in $B\to K^*\gamma$}

\begin{figure}
\includegraphics{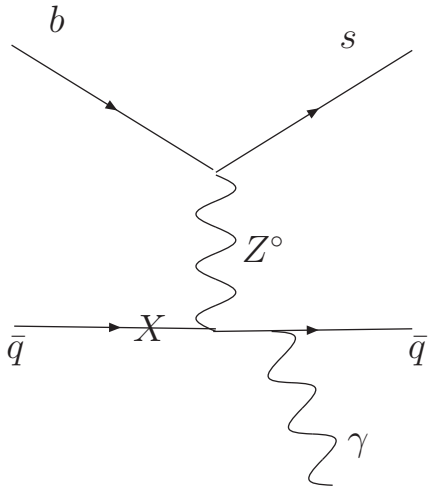}
\caption{\label{fig1} Tree level contribution to non-spectator processes in $B\to K^* \gamma$.  Cross represents
the alternative coupling of the emitted photon.}
\end{figure}

The non-vanishing FCNC at the tree level leads to an additional contributing Feynmann diagram which is
illustrated in Fig. \ref{fig1}.  The amplitude for $b\bar q\to s\bar q$ transition via $Z^0$ exchange in the VQM
can be written as\cite{ahmadynagashima}:
\begin{eqnarray}
A^{VQM}&=&\frac{ig}{2\cos (\theta )}\left(-\frac{1}{2}U^{sb}\right)\bar s\gamma^\mu (1-\gamma_5)b\times\frac{1}{M_Z^2}
\nonumber \\
&&\frac{ig}{2\cos (\theta )}\left[ (I^q_W-Q_q{\sin^2{\theta }})\bar q\gamma_\mu
(1-\gamma_5)q-Q_q{\sin^2{\theta}}\bar q\gamma_\mu (1+\gamma_5)q\right]\;\; ,\label{vqmamp}
\end{eqnarray}
where $U^{sb}=(V^\dagger V)^{sb}$ is a measure of the non-unitarity of the extended quark mixing matrix and
$I^q_W$ is the third component of the weak isospin of quark flavor $q$.  One can then write (\ref{vqmamp}) in
terms of the effective operators $O_3$ and $O_5$ which are defined as:
\begin{eqnarray}
O_3&=&\bar s_\alpha  \gamma^\mu (1-\gamma_5)b_\alpha \bar q_\beta\gamma_\mu (1-\gamma_5)q_\beta \;\; ,\nonumber \\
O_5&=& \bar s_\alpha  \gamma^\mu (1-\gamma_5)b_\alpha \bar q_\beta\gamma_\mu (1+\gamma_5)q_\beta \;\; .
\label{operators}
\end{eqnarray}
and therefore the contribution of the extra vector quarks results in additional terms in the Wilson coefficients
$C_3$ and $C_5$ to the leading order in the strong coupling $\alpha_s$.
\begin{eqnarray}
C_3^{VQM}&=&\frac{U^{sb}}{V_{tb}V_{ts}^*}(I_W^q-Q_q\sin^2{\theta})=
\frac{U^{sb}}{V_{tb}V_{ts}^*}\displaystyle\left\{^{\displaystyle
1/2-2/3\sin^2{\theta}
=0.35\ldots q=up}_{\displaystyle -1/2+1/3\sin^2{\theta}=-0.42\ldots q=down}\right. \;\; ,\nonumber \\
C_5^{VQM}&=&-\frac{U^{sb}}{V_{tb}V_{ts}^*}Q_q\sin^2{\theta}=
\frac{U^{sb}}{V_{tb}V_{ts}^*}\displaystyle\left\{^{\displaystyle
-2/3\sin^2{\theta}=-0.15\ldots q=up}_{\displaystyle 1/3\sin^2{\theta}=0.08\ldots q=down}\right. \;\; .
\label{wilsoncoef}
\end{eqnarray}
With the upper bound $\vert U^{sb}\vert\lesssim 10^{-3}$ coming from the rare B decays \cite{ahmadynagashima},
the additional contribution due to the tree level FCNC could be comparable to the SM value of these coefficients
at $\mu=m_b$, i.e. $C_3=0.014$ and $C_5=-0.041$.

Here an explanation is in order.  Strictly speaking, one should include the extra terms given in eq.
(\ref{wilsoncoef}), which are proportional to the electric charge of the light quark, in the electroweak penguin
operators $O_{7\ldots 10}$\cite{hiller}.  However, since, as far as nonspectator effects to the leading order of
$\alpha_s$ are concerned, one can ignore these operators within SM, we prefer to write the additional
VQM-generated contributions in terms of the dominant QCD penguin operators.  In any case, our results do not
change had we followed the strict formulation of the problem.

Following the method of Ref. \cite{kn}, one can write the nonspectator contributions to $B\to K^*\gamma$
amplitude as $A_q=b_qA_{lead}$, where $q$ is the flavor of the light anti-quark in the B meson and $A_{lead}$ is
the leading spectator amplitude.  To leading order in the strong coupling constant $\alpha_s$, the main
contribution to $B\to K^*\gamma$ is from the electromagnetic penguin operator $O_7$ and the factorizable
amplitude $A_{lead}$ is proportional to the form factor $T_1^{B\to K^*}$ which parameterizes the hadronic matrix
element of this operator to the leading order in $\Lambda_{QCD}/m_b$. $b_q$ is the parameter that depends on the
flavor of the spectator and, in fact, this parameterization leads to a simple expression for the isospin
asymmetry (as defined by eq. (\ref{isospinasym})) in terms of $b_q$:
\begin{equation}
\Delta_{0-}=\Re (b_{\bar d}-b_{\bar u})\;\; ,\label{asymb}
\end{equation}
Using the expression for $b_q$ which is derived within the QCD factorization method in Ref \cite{kn} , we obtain
the contribution of vector quarks to the isospin asymmetry as follows:
\begin{equation}
\Delta_{0-}^{VQM}=\Re\left(\frac{4\pi^2f_B}{N_cm_bT_1^{B\to
K^*}a_7^c}\frac{U^{sb}}{V_{tb}V_{ts}^*}\left[-0.22\frac{f_{K^*}^{\bot}F_{\bot}}{m_b}-0.28\frac{f_{K^*}m_{K^*}}
{6\lambda_Bm_B}\right]\right)\;\;
.\label{asymvqm}
\end{equation}
The numerical input for the parameters of eq. (\ref{asymvqm}) are tabulated in Table \ref{table1}, which results
in an isospin asymmetry due to the extra generation of quarks of the form:
\begin{equation}
\Delta_{0-}^{VQM}=-0.08\left\vert\frac{U^{sb}}{V_{tb}V_{ts}^*}\right\vert\cos{(\theta +\phi_s)}\;\; .
\label{asymnum}
\end{equation}
In the above formula, $\theta$ is the weak phase (CP odd) of the ratio $\frac{U^{sb}}{V_{tb}V_{ts}^*}$ in the
extended $4\times 4$ quark mixing matrix.  In a particular parametrization of the mixing matrix where $V_{tb}$
and $V_{ts}$ are taken to be real as in SM, $\theta$ is the phase of the nonunitarity parameter $U^{sb}$.  On
the other hand, $\phi_s$ is the strong phase (CP even) entering in eq. (\ref{asymvqm}). For example, the
imaginary part of the effective Wilson coefficient $a_7^c$ can be one possible source of this latter phase. We
would like to point out that the extra contribution due to vector-like quarks to $\Delta_{0+}$, i.e.
\begin{equation}
\Delta_{0+}^{VQM}=-0.08\left\vert\frac{U^{sb}}{V_{tb}V_{ts}^*}\right\vert\cos{(\theta -\phi_s)}\;\; ,
\label{asymnum2}
\end{equation}
is expected to be different from eq. (\ref{asymnum}) if $\phi_s$ is appreciable. It is interesting to see if the difference between $\Delta_{0-}$ and
$\Delta_{0+}$, as reflected in eqs. (\ref{babar}) and (\ref{belle}), will persist in the future measurements of the isospin asymmetry.  One could explain this
within the vector quark model if $\vert U^{sb}\vert$ happens to be around its upper allowed limit, i.e. a few times $10^{-3}$, with $\theta\sim\phi_s\sim\pi /4$.
 In case that the strong phase $\phi_s$ is negligible (clearly this is the case if the phase of $a_7^c$ is the only CP even phase in this transition), the isospin
asymmetry due to an extra generation of vector-like quarks is sensitive only to the magnitude and phase of $U^{sb}$, and therefore, with more precise
experimental data becoming available in the future, this observable could serve to impose a stringent constrain on the important nonunitarity parameter of the
vector quark model.
\begin{table}
\caption{\label{table1}The numerical values of the parameters in eqn. (\ref{asymvqm}).  }
\begin{ruledtabular}
\begin{tabular}{ccccccccc}
$T_1$&$m_b$&$\lambda_B$&$f_{K^*}$&$f_{K^*}^{\bot}$&$m_B$&$m_{K^*}$&$F_{\bot}$&$a_7^c$\\
\hline
0.32& 4.2 GeV & 0.35 GeV&0.226 GeV&0.175 GeV&5.28 GeV&0.892 GeV&1.21&-0.41-0.03i \cite{bosch}\\
\end{tabular}
\end{ruledtabular}
\end{table}

\section{\label{section3}Direct CP violation within the VQM}
In case that the strong phase $\phi_s$ in eq. (\ref{asymnum}) is significant, one could look into another
important observable in the $B\to K^*\gamma$ transition, i.e. direct CP violation , which in combination with
isospin asymmetry help to constrain $U^{sb}$.  Direct CP asymmetry, which is defined as:
\begin{equation}
A_{CP}=\frac{\Gamma (B^+\to K^{*+}\gamma )-\Gamma (B^-\to K^{*-}\gamma )}{\Gamma (B^+\to K^{*+}\gamma )+\Gamma
(B^-\to K^{*-}\gamma )}\;\; ,\label{cpasym}
\end{equation}
is nonzero if at least two different diagrams with non-identical weak and strong phases contribute to the decay
process.  In other words, for $B\to K^*\gamma$ transition, we should expect non-vanishing $A_{CP}$ from
nonspectator processes if $b_q$ happens to include both strong as well as weak phases.  In this case, eq.
(\ref{cpasym}) can be written as:
\begin{equation}
A_{CP}=\Re (b_u-b_{\bar u})\;\; .\label{cpbu}
\end{equation}
The SM prediction for direct CP violation in $B\to K^*\gamma$ is vanishingly small
 within the theoretical error.  For example, using the
 perturbative QCD method, it is calculated to be $A_{CP}=(0.62\pm 0.13)\times 10^{-2}$\cite{kms}.  Therefore,
 any significant CP asymmetry is an indication of new physics.
In our case, the contribution of extra vector-like quarks to eq. (\ref{cpbu}) within the QCD factorization
method is obtained as follows:
\begin{equation}
A^{VQM}_{CP}=\left(\frac{16\pi^2f_B}{N_cm_bT_1^{B\to K^*}}\left\vert\frac{U^{sb}}{a_7^c
V_{tb}V_{ts}^*}\right\vert\left[0.15\frac{f_{K^*}^{\bot}F_{\bot}}{m_b}+0.35\frac{f_{K^*}m_{K^*}}{6\lambda_Bm_B}
\right]\right)\sin{\theta}\sin{\phi_s}\;\;
.\label{cpasymvqm}
\end{equation}
Inserting the values given in Table \ref{table1} for the parameters in the above formula leads to a simple
expression of the additional direct CP violation in terms of the nonunitarity parameter, its CP odd phase and
the strong phase:
\begin{equation}
A_{CP}^{VQM}=0.27\left\vert\frac{U^{sb}}{V_{tb}V_{ts}^*}\right\vert\sin{\theta}\sin{\phi_s}\;\; .
\label{cpasymnum}
\end{equation}
 The available experimental data on this asymmetry, i.e. $A_{CP}=0.007\pm 0.074\pm 0.017$ \cite{babar}
 has large errors and is consistent with zero.  The combination of isospin (eqns. (\ref{asymnum}) and \ref{asymnum2})) and direct CP (eq. (\ref{cpasymnum}))
 asymmetries due to vector quarks leads to the following prediction: if the difference between $\Delta_{0-}$ and $\Delta_{0+}$ is mainly due to extra vector quarks
 then one expects a direct CP asymmetry in $B\to K^*\gamma$ of around $6-7\%$.  It will be exciting to see if the more accurate experimental measurements in the
 future will result in a significant shift of the central value of $A_{CP}$.

\begin{acknowledgments}
M.A.'s research is partially funded by a discovery grant from NSERC.
\end{acknowledgments}

\end{document}